\documentclass[article, noheadings, nofooter, nojss]{jss}\usepackage[]{graphicx}\usepackage[]{color}
\makeatletter
\def\maxwidth{ %
  \ifdim\Gin@nat@width>\linewidth
    \linewidth
  \else
    \Gin@nat@width
  \fi
}
\makeatother

\usepackage{color}
\definecolor{green}{rgb}{0, 0.5, 0.0}
\definecolor{brown}{rgb}{0.59, 0.29, 0.0}
\usepackage{alltt}
\usepackage{amsmath}
\usepackage{amssymb}
\usepackage{natbib}
\usepackage{bm}

\author{Benjamin S\"afken\\Georg-August Universit\"at G\"ottingen \And 
		David R\"ugamer\\Ludwig-Maximilans-Universit\"at M\"unchen \AND
        Thomas Kneib\\Georg-August Universit\"at G\"ottingen \And
		Sonja Greven\\Ludwig-Maximilans-Universit\"at M\"unchen}
		
\title{Conditional Model Selection in Mixed-Effects Models with \pkg{cAIC4}}

\Plainauthor{Benjamin S\"afken, David R\"ugamer, Thomas Kneib, Sonja Greven} 
\Plaintitle{Conditional Model Selection in Mixed-effects Models with lme4} 
\Shorttitle{\pkg{cAIC4}: Conditional AIC for \pkg{lme4}} 

\Abstract{
Model selection in mixed models based on the conditional distribution is 
appropriate for many practical applications and has been a focus of recent 
statistical research. In this paper we introduce the \textsf{R}-package 
\pkg{cAIC4} that allows for the computation of the conditional Akaike 
Information Criterion (cAIC). Computation of the conditional AIC needs to take
into account the uncertainty of the random effects variance and is therefore not 
straightforward. We introduce a fast and stable implementation for the 
calculation of 
the cAIC for 
linear mixed models estimated with \pkg{lme4} 
and additive mixed models estimated with \pkg{gamm4}
. 
Furthermore, \pkg{cAIC4} offers a stepwise function that allows for a fully automated 
stepwise selection scheme for mixed models based on the conditional AIC. Examples of many 
possible applications are presented to illustrate the practical impact and easy 
handling of the package.\\
}
\Plainkeywords{conditional AIC, lme4, mixed-effects models, penalized splines} 

\Address{
  Benjamin S\"afken\\
  RTG 1644: Scaling Problems in Statistics\\
  Universit\"at G\"ottingen\\
  37073 G\"ottingen, Germany\\
  E-mail: \email{bsaefke@uni-goettingen.de}\\
  URL: \url{http://www.uni-goettingen.de/de/322166.html}\\[.3cm]
  David R\"ugamer\\
  Ludwig-Maximilans-Universit\"at M\"unchen\\
  E-mail: \email{david.ruegamer@stat.uni-muenchen.de}\\[.3cm]
  Sonja Greven\\
  Ludwig-Maximilans-Universit\"at M\"unchen\\
  E-mail: \email{sonja.greven@stat.uni-muenchen.de}\\[.3cm]
  Thomas Kneib\\
  Chair of Statistics\\
  Universit\"at G\"ottingen\\
  37073 G\"ottingen, Germany\\
  E-mail: \email{tkneib@uni-goettingen.de}\\
  URL: \url{http://www.uni-goettingen.de/de/264255.html}\\ 
}

\IfFileExists{upquote.sty}{\usepackage{upquote}}{}
\begin{document}


\section{Introduction}

The linear mixed model is a flexible and broadly applicable statistical model.
It is naturally used for analysing longitudinal or clustered data. Furthermore,
any regularized regression model incorporating a quadratic penalty can be written in terms
of a mixed model. This incorporates smoothing spline models, spatial models and 
more general additive models \citep{Wood.2017}. Thus efficient and reliable estimation of such
models is of major interest for applied statisticians. The package 
\pkg{lme4} for the statistical computing software \textsf{R} \citep{R.2016} offers 
such an exceptionally fast and generic implementation for 
mixed models \citep[see][]{lme4.2015}. The package has a modular framework allowing for the profile 
restricted maximum likelihood (REML) criterion as a function of the model 
parameters to be optimized using any constrained optimization function in \textsf{R} and 
uses rapid techniques for solving penalized least squares problems based on
sparse matrix methods.\\
The fact that mixed models are widely used popular statistical tools make model 
selection an indispensable necessity. Consequently research regarding model 
choice, variable selection and hypothesis testing in mixed models has flourished 
in recent years.\\
Hypothesis testing on random effects is well established, although for likelihood 
ratio tests boundary issues arise \citep{Crainiceanu.2004,Greven.2008,Wood.2013b}.
In model selection for mixed models using the Akaike information criterion
\citep[AIC][]{Akaike.1973}, \citet{Vaida.2005} suggest to use different criteria
depending on the focus of the underlying research question. They make a 
distinction between questions with a focus on the population and on clusters, respectively. 
For the latter, they introduce a conditional AIC accounting for the shrinkage in the
random effects. Based on this conditional AIC, \cite{Liang.2008} propose a
criterion that corrects for the estimation uncertainty of the random effects 
variance parameters based on a numerical approximation. \cite{Greven.2010} show
that ignoring this estimation uncertainty induces a bias and derive
an analytical representation for the conditional AIC.\\
For certain generalized mixed models, analytical representations of the conditional AIC exist, 
for instance for Poisson responses \citep[see][]{Lian.2012}. Although there
is no general unbiased criterion in analytical form for all exponential family distributions 
as argued in \cite{Saefken.2014}, bootstrap-based methods can often be applied as 
we will show for those in 
presented in \cite{Efron.2004}.
An asymptotic criterion for a wider class of distributions is described in 
\cite{Wood.2016}.\\ 
In this paper, we describe an add-on package to \pkg{lme4} that facilitates 
model selection based on the conditional AIC and 
illustrates it with several examples. For the conditional AIC proposed by 
\cite{Greven.2010} for linear mxied models, 
the computation of the criterion is not as simple as it 
is for other common AIC criteria. This article focuses on techniques for fast 
and stable computation of the conditional AIC in mixed models estimated with 
\pkg{lme4}, as they are implemented in the \textsf{R}-package 
\pkg{cAIC4}. The amount of possible models increases substantially with the 
\textsf{R}-package \pkg{gamm4} \citep[see][]{gamm4.2016} allowing for the estimation of a wide class of 
models with quadratic penalty such as spline smoothing and additive models. The
presented conditional AIC applies to any of these models.\\
In addition to translating the findings of \cite{Greven.2010} to the model 
formulations used in \cite{lme4.2015}, we present the implementation of conditional AICs 
proposed for non-Gaussian settings in \cite{Saefken.2014} and 
as we propose based on 
\cite{Efron.2004}. 
With these results, a new scheme for stepwise conditional variable
selection in mixed models is introduced. This allows for fully automatic choice 
of fixed and random effects based on the optimal conditional AIC. All methods 
are accompanied by examples, mainly taken from \pkg{lme4}, see 
\cite{lme4.2015}. The rest of this paper is structured as follows:\\
In Section 2 the mixed
model formulations are introduced based on one example with random intercepts and 
random slopes and a second example on penalised spline smoothing. The 
conditional AIC for Gaussian, Poisson and Bernoulli responses is introduced in 
Section 3. Section 4 gives a hands-on introduction to \pkg{cAIC4} with
specific examples for the \code{sleepstudy} and the \code{grouseticks} data 
from \pkg{lme4}. The new scheme for stepwise conditional variable selection in 
mixed models is presented in Section 5 and applied to the \code{Pastes} data
set. After the conclusion in Section 6, part A of the appendix describes how
\pkg{cAIC4} automatically deals with boundary issues. Furthermore the 
underlying code for the rapid computation of the conditional AIC is presented in 
part B of the appendix.\\

\section{The mixed model} \label{sec:MM}

In a linear mixed model, the conditional distribution of the response $\boldsymbol{y}$
given the random effects $\boldsymbol{u}$ has the form

\begin{equation}
  \boldsymbol{y}|\boldsymbol{u} \sim 
  \mathcal{N}\left(\boldsymbol{X}\boldsymbol{\beta} + \boldsymbol{Z}\boldsymbol{u}, \sigma^2\boldsymbol{I}_n\right), \label{eq:lmm}
\end{equation}

where $\boldsymbol{y}$ is the $n$-dimensional vector of responses, $\boldsymbol{\beta}$ is the
$p$-dimensional vector of fixed effects and $\boldsymbol{u}$ is the $q$-dimensional vector of random effects. The matrices
$\boldsymbol{X}$ and $\boldsymbol{Z}$ are the $(n \times p)$ and $(n \times q)$
design matrices for fixed and random effects, respectively, and $\sigma^2$ refers to the variance of the error terms.\\
The unconditional distribution of the random effects $\boldsymbol{u}$ is assumed to be a 
multivariate Gaussian with mean $\boldsymbol{0}$ and positive semidefinite $(q 
\times q)$ covariance matrix $\boldsymbol{D}_{\boldsymbol{\theta}}$, i.e., 

\[
  \boldsymbol{u} \sim \mathcal{N}\left(\boldsymbol{0}, \boldsymbol{D}_{\boldsymbol{\theta}}\right).
\]

The symmetric covariance matrix $\boldsymbol{D}_{\boldsymbol{\theta}}$ depends on the covariance 
parameters $\boldsymbol{\theta}$ and may be decomposed as

\begin{equation}
  \boldsymbol{D}_{\boldsymbol{\theta}} = \sigma^2
  \boldsymbol{\Lambda}_{\boldsymbol{\theta}}\boldsymbol{\Lambda}_{\boldsymbol{\theta}}^t, \label{eq:DMatrix}
\end{equation}

with the lower triangular covariance factor 
$\boldsymbol{\Lambda}_{\boldsymbol{\theta}}$ and the variance parameter 
$\sigma^2$ of the conditional response distribution.
In analogy to generalized linear models, the generalized linear mixed model 
extends the distributional assumption in (\ref{eq:lmm}) to a distribution 
$\mathcal{F}$ from the exponential family,

\[
  \boldsymbol{y}|\boldsymbol{u} \sim \mathcal{F} (\boldsymbol{\mu}, \phi)
\]

where $\phi$ is a scale parameter and the mean has the form

\begin{equation}
  \bm{\mu} = \mathbb{E}(\boldsymbol{y}|\boldsymbol{u}) =
h\left(\boldsymbol{X}\boldsymbol{\beta} + \boldsymbol{Z}\boldsymbol{u}\right),
\label{eq:glmmLink}
\end{equation}

with $h$ being the response function applied componentwise and natural parameter
$\boldsymbol{\eta} = h^{-1}\left(\boldsymbol{\mu}\right)$. As the hereinafter 
presented results are limited to the Poisson and binomial distributions we can 
assume $\phi = 1$. The symmetric covariance matrix in (\ref{eq:DMatrix}) then is 
the same as for Gaussian responses except that $\sigma^2$ is omitted, i.e., 
$\boldsymbol{D}_{\boldsymbol{\theta}} = 
\boldsymbol{\Lambda}_{\boldsymbol{\theta}}\boldsymbol{\Lambda}_{\boldsymbol{\theta}}^t$.\\

The given conditional formulations of (generalized) linear mixed models imply 
marginal models, which can (conceptually) be obtained by integrating the random 
effects out of the joint distribution of $\bm{y}$ and $\bm{u}$, i.e., 
$$
f(\bm{y}) = \int f(\bm{y}\mid\bm{u}) f(\bm{u}) d\bm{u}.
$$ 
However, there is typically no closed form solution for this integral.
While the marginal model formulation is usually
used for estimation, an analytic representation of $f(\bm{y})$ is only available
for the linear mixed model (\ref{eq:lmm}). The marginal distribution $f(\bm{y})$
for Gaussian responses $\bm{y}$ is given by

$$
\bm{y} \sim \mathcal{N}
\left(\boldsymbol{X}\boldsymbol{\beta}, \sigma^2 \left( \boldsymbol{I}_n +
\boldsymbol{Z}\boldsymbol{\Lambda}_{\boldsymbol{\theta}} 
\boldsymbol{\Lambda}_{\boldsymbol{\theta}}^t\boldsymbol{Z}^t\right)\right).
$$
  
Further extensions of linear mixed models can be obtained by, for example, 
relaxing the assumption $\text{Cov}(\bm{y}|\bm{u}) =  \sigma^2\boldsymbol{I}_n$. 


\subsection*{Example I: Random intercepts and random slopes}

Some special cases of mixed models are commonly used in applications, including
the random intercept model and the random slope model. In the
random intercept model, the responses differ in an individual- or
cluster-specific intercept for $m$ individuals or clusters. In this case the
individual-specific intercept is modeled as random effect $\bm{u} = (u_{1,1},
u_{1,2}, \ldots, u_{1,m})$, yielding the (generalized) linear mixed model 

\[
\mathbb{E}(y_{ij}|u_{1,i}) = h(\bm{x}_{ij} \bm{\beta} + u_{1,i}),\quad
u_{1,i} \overset{iid}{\sim} \mathcal{N}(0,\tau_0^2 \bm{I}_m)
\]

for the $j$-th observation from an individual or cluster $i$.\\

Whereas for the random intercept model all covariates modeled with fixed effects 
are assumed to have the same influence on the response variable across individuals, the random
slope model is suitable when an independent variable $x_s$ is assumed to have an
individual-specific effect on the dependent variable. The random intercept model
is extended to $$\mathbb{E}(y_{ij}|\bm{u}_i) = h(\bm{x}_{ij} \bm{\beta} +
u_{1,i} + x_{s,ij} u_{2,i}),$$ where $~u_{2,i}~$ is the individual-specific
slope, which can be regarded as the deviation from the population slope $~\beta_s~$
corresponding to the $s$-th covariate $~x_{s,ij}~$ in
$~\bm{x}_{ij}$. In most cases, there is no reason to suppose $u_{1,i}$ and $u_{2,i}$ to be uncorrelated and the distributional
assumption thus is

\begin{equation}
\begin{pmatrix}
u_{1,i} \\
u_{2,i} 
\end{pmatrix} \sim \mathcal{N} \left(
 \begin{pmatrix} 0 \\ 0 \end{pmatrix}
 , 
 \begin{pmatrix}
 \tau_1^2 &
 \tau_{12} \\
 \tau_{21} &
 \tau_2^2
 \end{pmatrix} \right).
\end{equation}

\vspace*{1cm}

\subsection*{Example II: Penalised spline smoothing}

In addition to many possibilities to extend these simple random effect models, linear
mixed models can also be utilized to fit semi-parametric regression models \citep[see, e.g.,][]{Ruppert.2003}. 
For univariate smoothing, consider the model 

\begin{equation}
\mathbb{E}(y_i) = f(x_i), \label{eq:semipar}
\end{equation}

for $i=1,\ldots,n$, where $f(\cdot)$ is a deterministic function of the covariate
$x_i$, which shall be approximated using splines. For illustrative purposes, we consider
the truncated polynomial basis representation

\begin{equation}
f(x) = \sum_{j=0}^g \beta_j x^j + \sum_{j=1}^k u_j (x - \kappa_j)_{+}^g, \label{eq:tpseries}
\end{equation}
in the following, 
where $\kappa_1 < \ldots < \kappa_k$ are $k \in \mathbb{N}$ knots, partitioning 
the domain of $x$, $g \in \mathbb{N}$ and 

\begin{equation}
(z)_{+}^g = z^g \cdot I(z>0) = 
\begin{cases}
   z^g & \text{if } z > 0 \\
   0       & \text{if } z \leq 0
  \end{cases}.
\end{equation}

As the truncated part $u_j (x-\kappa_j)^g_{+}$ is non-zero for $x > \kappa_j$,
$u_j$ can be seen as a gradient change of the two consecutive function segments 
defined on $(\kappa_{j-1},\kappa_j]$ and $(\kappa_j,\kappa_{j+1}]$. In order to 
estimate $\beta_j, j=0,\ldots,g$ and $u_j, j=1,\ldots,k$, the method of ordinary
least squares (OLS) could in principle be applied. 
In most cases, however, this yields a rather 
rough estimate of $f$ for suitably large $k$ as the gradient changes of functions 
segments have a large impact. Therefore estimation methods for linear mixed models can be 
utilized in order to obtain a smooth function. Representing the untruncated 
polynomial part in (\ref{eq:tpseries}) as the fixed effects and
$\sum_{j=1}^k u_j (x-\kappa_j)^g_{+}$ as the random effects part, the well known
shrinkage effect of mixed models is transferred to the estimation of the $u_j$s,
shrinking the changes in the gradient of the fitted polynomials. 
The random effects assumption corresponds to a quadratic penalty on the $u_j$, with 
the smoothing parameter estimated from the data.

This approach also works analogously for various other basis functions including 
the frequently used B-spline basis \citep[see, e.g.,][]{Fahrmeir.2013}. Moreover, a
rich variety of models that can be represented as reduced rank basis smoothers with 
quadratic penalties allow for this kind of representation. The estimation via
\pkg{lme4} can be employed by the use of \pkg{gamm4}. For an overview of 
possible model components see \cite{Wood.2017}. An example is also given in 
Section 5.

\section{The conditional AIC}

\subsection*{The Akaike Information Criterion}

Originally proposed by Hirotogu Akaike \citep{Akaike.1973}
as An Information Criterion (AIC), the AIC was one of the
first model selection approaches to attract special attention among users of
statistics. In some way, the AIC extends the maximum likelihood paradigm by
making available a framework, in which both parameter estimation and model
selection can be accomplished. The principle idea of the AIC can be traced back
to the Kullback-Leibler distance \citep[KLD][]{Kullback.1951}, which can be used to measure
the distance between a true (but normally unknown) density $g(\bm{y})$ and a 
parametric model $f(\bm{y}\mid\bm{\nu})$. The unknown parameters $\bm{\nu}$ are 
commonly estimated by their maximum likelihood estimator 
$\hat{\bm{\nu}}(\bm{y})$. As minimizing the expected Kullback-Leibler distance 
is equivalent to minimizing the so called Akaike Information 

\begin{equation}
\text{AI} = - 2 \, \mathbb{E}_{g(\bm{y})} \mathbb{E}_{g(\tilde{\bm{y}})} \log f(\tilde{\bm{y}}\mid\hat{\bm{\nu}}(\bm{y})),
\label{eq:AI}
\end{equation}

with $\tilde{\bm{y}}$ a set of independent new observations from $g$, minus twice the maximized log-likelihood $\log f(\bm{y} \mid \hat{\bm{\nu}} (\bm{y}))$ as a natural measure of goodness-of-fit is an obvious
estimator of the AI. However, this approach induces a bias as the maximized
log-likelihood only depends on $\bm{y}$ whereas (\ref{eq:AI}) is
defined as a predictive measure of two independent replications $\tilde{\bm{y}}$
and $\bm{y}$ from the same underlying distribution. Therefore the bias 
correction is defined by

\begin{equation}
\text{BC} = 2\left(\mathbb{E}_{g(\bm{y})} \log f(\bm{y} \mid \hat{\bm{\nu}} (\bm{y})) -  
\, \mathbb{E}_{g(\bm{y})} \mathbb{E}_{g(\tilde{\bm{y}})} \log f(\tilde{\bm{y}}\mid\hat{\bm{\nu}}(\bm{y}))\right).
\label{eq:BC}
\end{equation}

Akaike derived the bias correction, which under certain regularity conditions 
can be estimated asymptotically by two times the dimension of $\bm{\nu}$. This yields the 
well-known AI estimator 

\[
\text{AIC}(\bm{y}) = -2 \log f(\bm{y} \mid \hat{\bm{\nu}}(\bm{y})) +
2\, \text{dim}(\bm{\nu}).
\] 

Hence, as the statistical model $f(\cdot|\bm{\nu})$
with the smallest AI aims at finding the model which is closest to the true
model, the AIC can be seen as a relative measure of goodness-of-fit for
different models of one model class. Notice that the bias correction is 
equivalent to the (effective) degrees of freedom and the covariance penalty, 
see \cite{Efron.2004}.


\subsection*{The marginal and the conditional perspective on the AIC}

Adopting this principle for the class of mixed models to select amongst 
different random effects is not straightforward. First of all, the question 
arises on the basis of which likelihood to define this AIC. For the class of 
mixed models, two common criteria exist, namely the marginal AIC (mAIC) based on 
the marginal log-likelihood and the conditional AIC (cAIC) based on the
conditional log-likelihood. The justification of both approaches therefore
corresponds to the purpose of the marginal and the conditional mixed model 
perspective, respectively. Depending on the question of interest, 
the intention of both perspectives differs, as for 
example described in \citet{Vaida.2005} or \citet{Greven.2010}. 

The marginal perspective of mixed models is suitable when the main interest 
is to model fixed population effects with a 
reasonable correlation structure. The conditional perspective, by contrast, can be 
used to make statements based on the fit of the predicted random effects. In longitudinal 
studies, for example, the latter point of view seems to be more appropriate if 
the focus is on subject- or cluster-specific random effects. Another crucial 
difference in both approaches lies in the model's use for prediction. On the one 
hand, the marginal model seems to be more plausible if the outcome for new 
observations comes from new individuals or clusters, i.e., observations 
having new random effects. The conditional model on the other hand is 
recommended if predictions are based on the same individuals or clusters, thereby 
predicting on the basis of already modeled random effects. 

The corresponding AI criteria have closely related intentions. The conditional AIC 
estimates the optimism of the estimated log-likelihood for a new data set
$\tilde{\bm{y}}$ by leaving the random effects unchanged. This can be understood 
as a predictive measure based on a new data set originating from the 
same clusters or individuals as $\bm{y}$. On the contrary, the marginal approach 
evaluates the log-likelihood using a new predictive data set $\tilde{\bm{y}}$,
which is not necessarily associated with the cluster(s) or individual(s) of 
$\bm{y}$. 

In particular for the use of mixed models in penalized spline smoothing, the 
cAIC usually represents a more plausible choice. As demonstrated in Example II 
of Section~\ref{sec:MM}, the representation of penalized spline smoothing via
mixed models divides certain parts of the spline basis into fixed and random 
effects. Using the marginal perspective in Example II, predictions would 
therefore be based only on the polynomial coefficients of $f$. If the fitted 
non-linear function is believed to represent a general relationship of $x$ and 
$y$, predictions as well as the predictive measure in terms of the Akaike 
Information, however, make more sense if the truncated parts of the basis are 
also taken into account.\\


\citet{Vaida.2005} proposed the cAIC, an estimator of 
the conditional Akaike Information 

\begin{equation}
\text{cAI} = -2\,\mathbb{E}_{g(\bm{y},\bm{u})}
\mathbb{E}_{g(\tilde{\bm{y}}|\bm{u})} \log f(\tilde{\bm{y}} \mid \hat{\bm{\nu}}
(\bm{y}), \hat{\bm{u}}(\bm{y}))
\label{cAI}
\end{equation}

as an alternative to the mAIC, where $\bm{\nu}$ includes the fixed effects and 
covariance parameters $\bm{\theta}$. The cAIC may be more appropriate when the AIC is 
used for the selection of random effects. In addition, \citet{Greven.2010} 
investigated the difference of both criteria 
from a mathematical point of view. Since the mAIC is intended for the use in 
settings where the observations are independent and the $k$-dimensional 
parameter space $\bm{\mathcal{V}}_k$ can be transformed to $\mathbb{R}^k$, the 
corresponding bias correction $2\, \text{dim}(\bm{\nu})$ is biased for mixed 
models for which these conditions do not apply. In particular, Greven and Kneib 
showed that the mAIC leads to a preference for the selection of smaller models 
without random effects. 

\subsection*{Conditional AIC for Gaussian responses}

Depending on the distribution of $\bm{y}$, different
bias corrections of the maximized conditional log-likelihood exist to obtain the cAIC.
For the Gaussian case, \citet{Liang.2008} 
derive a corrected version of the initially proposed cAIC by \citet{Vaida.2005} 
for known error variance, taking into account the estimation of the covariance 
parameters $\bm{\theta}$: 

\begin{equation}
\text{cAIC}(\bm{y}) = -2 \log f({\bm{y}} \mid \hat{\bm{\nu}} (\bm{y}),
\hat{\bm{u}}(\bm{y})) + 2\, \text{tr}\left(\frac{\partial \hat{\bm{y}}}{\partial \bm{y}}\right).  \label{eq:cAIC}
\end{equation}

Evaluating the bias correction $\text{BC} = 2\, \text{tr}(\frac{\partial \hat{\bm{y}}}{\partial \bm{y}})$ 
in expression (\ref{eq:cAIC}) via numerical approximation, or a similar formula for unknown error 
variance, is however computationally expensive. 
\citet{Greven.2010} develop an analytic version of the corrected cAIC making the 
calculation of the corrected cAIC feasible. We adapt their efficient 
implementation originally written for \code{lme}-objects (returned by the \code{nlme} package) and reimplement their
algorithm for \code{lmerMod}-objects (returned by \code{lme4}). A more detailed
description on the calculation of several terms in the proposed formula of 
\citet{Greven.2010} is given in Appendix~\ref{sec:compMat}. Furthermore, a
partition of the parameter space is needed in order to account for potential 
parameters on the boundary of the parameter space, as presented in Theorem 3 in
\cite{Greven.2010}. This process can be very unwieldy. Therefore, a fully 
automated correction algorithm is implemented in \pkg{cAIC4} and presented in 
Appendix~\ref{sec:boundary}.\\\ 

\subsection*{Conditional AIC for Poisson responses}

As for the Gaussian case, 
note that for the Poisson and the binomial distribution the bias 
correction (\ref{eq:BC}) can be rewritten as twice the sum of the covariances between $\widehat{\eta}_i$ and $y_i$,

\begin{equation}
   BC = 
   2
   \sum_{i=1}^n \mathbb{E} \left( \widehat{\eta}_i \left( y_i - \mu_i \right) \right),
\label{eq:CovPen}
\end{equation}

with true but unobserved mean $\mu_i$ and the estimator of the natural parameter 
$\widehat{\boldsymbol{\eta}}$ depending on $\boldsymbol{y}$.
For the Poisson distribution an analytic reformulation of the bias correction 
term (\ref{eq:CovPen}) has to be utilized to make it analytically accessible as in \citet{Saefken.2014}.
Using results from \citet{Hudson.1978} and an identity due to \citet{Chen.1975}, the
bias correction (\ref{eq:CovPen}) for Poisson distributed responses can be 
reformulated to 

\begin{equation}
BC = 2 \sum_{i=1}^n  \mathbb{E} \left( y_i \left( \log \hat{\mu}_i(\bm{y}) -
\log \hat{\mu}_i(\bm{y}_{-i},y_i-1) \right) \right), \label{eq:cAICpo}
\end{equation}

for observations $i=1,\ldots,n$ and mean estimator $\hat{\mu}_i$. The $i$-th component of
$\bm{y}$ in $(\bm{y}_{-i}, y_i-1)$ is substituted by $y_i-1$ along with the
convention $y_i \log \hat{\mu}_i(\bm{y}_{-i},y_i-1) = 0$ if $y_i = 0$. 
The computational implementation of the cAIC in this case requires $n-d$ model 
fits, where $d$ corresponds to the number of Poisson responses being equal to 
zero (see Section~\ref{sec:cAIC_posBin} for details).
The resulting cAIC was first derived by \citet{Lian.2012}.

\subsection*{Conditional AIC for Bernoulli responses}

For binary responses there is no analytical representation for the bias 
correction (\ref{eq:CovPen}), see \citet{Saefken.2014}. Nevertheless a
bootstrap estimate for the bias correction 
can be based on 
\citet{Efron.2004}. 
The bias correction is equal to the 
sum over the covariances of the estimators of the natural parameter $\widehat{\eta}_i$ and the 
data $y_i$. To estimate this quantity, we could in principle draw a parametric bootstrap sample $\boldsymbol{z}_i$ of size B for the 
$i$-th data point - keeping all other observations fixed at their observed values - to estimate the $i$-th component 
$\mathbb{E} \left( \widehat{\eta}_i \left( y_i - \mu_i \right) \right)$ of the bias correction 
(\ref{eq:CovPen}) for binary responses 
by 

\begin{equation*}
    \ \frac{1}{B - 1} \sum_{j = 1}^B
   \hat{\eta}_i(z_{ij}) \left( z_{ij} - \overline{\boldsymbol{z}}_{i\cdot} \right)
   = \frac{B_1}{B - 1} \hat{\eta}_i(1)
   \left( 1 - \overline{\boldsymbol{z}}_{i\cdot} \right)
   + \frac{B_0}{B - 1} \hat{\eta}_i(0) 
   \left(- \overline{\boldsymbol{z}}_{i\cdot} \right),
\end{equation*}

where $B_0$ is the number of zeros in the bootstrap sample, $B_1$ is the 
number of ones in the bootstrap sample, $\hat{\eta}_i(1) = \log 
\left(\frac{\hat{\mu}_i(1)}{1-\hat{\mu}_i(1)}\right)$ is the estimated logit 
(the natural parameter) with $z_{ij} = 1$, $\hat{\eta}_i(0)= \log 
\left(\frac{\hat{\mu}_i(0)}{1-\hat{\mu}_i(0)}\right)$ is the estimated logit 
with $z_{ij}=0$ and $\overline{\boldsymbol{z}}_{i\cdot}$ is the mean of the 
bootstrap sample $\boldsymbol{z}_i$. Letting the number of bootstrap samples 
tend to infinity, i.e., $B \rightarrow \infty$ the mean of the bootstrap sample 
$\overline{\boldsymbol{z}}_{i\cdot} = \frac{1}{B}\sum_{j=1}^B z_{ij} = B_1/B$ (as well as $B_1 / (B-1)$ ) converges 
to the estimate from the data, which corresponds to the true mean in the bootstrap, 
$\hat{\mu}_i$
and therefore

\begin{align*} 
 \frac{B_1}{B - 1} \hat{\eta}_i(1)
   \left( 1 - \overline{\boldsymbol{z}}_{i\cdot} \right) -
   \frac{B_0}{B - 1} \hat{\eta}_i(0) 
   \left(\overline{\boldsymbol{z}}_{i\cdot} \right)
&\rightarrow\hat{\mu}_i \hat{\eta}_i(1)\left( 1 - \hat{\mu}_i \right)
 - \left(1 - \hat{\mu}_i\right) \hat{\eta}_i(0)\left(\hat{\mu}_i \right)\\
&= \hat{\mu}_i \left(1 - \hat{\mu}_i\right)
\left(\hat{\eta}_i(1) -\hat{\eta}_i(0)\right) \mbox{ for } B \rightarrow \infty.
\end{align*}

Since the bootstrap estimates are optimal if the number of bootstrap samples 
$B$ tends to infinity, this estimator can be seen as the optimal bootstrap 
estimator. The resulting estimator of the bias correction 

\begin{equation}
\widehat{BC} = 2 \sum_{i=1}^n \hat{\mu}_i \left(1 - \hat{\mu}_i\right)
\left(\hat{\eta}_i(1) -\hat{\eta}_i(0)\right)
\label{eq:BerBC}
\end{equation}

, which we use in the following, avoids a full bootstrap but requires $n$ model refits. 

\section{Introduction to cAIC4}

\subsection*{Example for linear mixed models}

An example that is often used in connection with the \textsf{R}-package 
\pkg{lme4} is the \code{sleepstudy} data from a study on the daytime 
performance changes of the reaction time during chronic sleep restriction, see 
\cite{Belenky.2003}. Eighteen volunteers
were only allowed to spend three hours of their daily time in bed for one week. 
The speed (mean and fastest 10\% of responses) and lapses (reaction times 
greater than 500 ms) on a psychomotor vigilance task where measured several 
times. The averages of the reaction times are saved as response variable 
\code{Reaction} in the data set. Each volunteer has an identifier 
\code{Subject}. Additionally the number of days of sleep restriction at each 
measurement is listed in the covariate \code{Days}. 

An example of how the \code{sleepstudy} data looks can be derived by the first 
13 of the 180 measurements it contains: 

\begin{Schunk}
\begin{Sinput}
R> sleepstudy[1:13,]
\end{Sinput}
\begin{Soutput}
   Reaction Days Subject
1  249.5600    0     308
2  258.7047    1     308
3  250.8006    2     308
4  321.4398    3     308
5  356.8519    4     308
6  414.6901    5     308
7  382.2038    6     308
8  290.1486    7     308
9  430.5853    8     308
10 466.3535    9     308
11 222.7339    0     309
12 205.2658    1     309
13 202.9778    2     309
\end{Soutput}
\end{Schunk}

Further insight into the data can be gained by a lattice plot, as presented in 
\cite{lme4.2015}. The average 
reaction times of each volunteer are plotted against the days of sleep 
restriction with the corresponding linear regression line. Such a plot can be 
found in Figure~\ref{fig:sleepPlot}.


\begin{Schunk}
\begin{figure}

{\centering \includegraphics[width=\maxwidth]{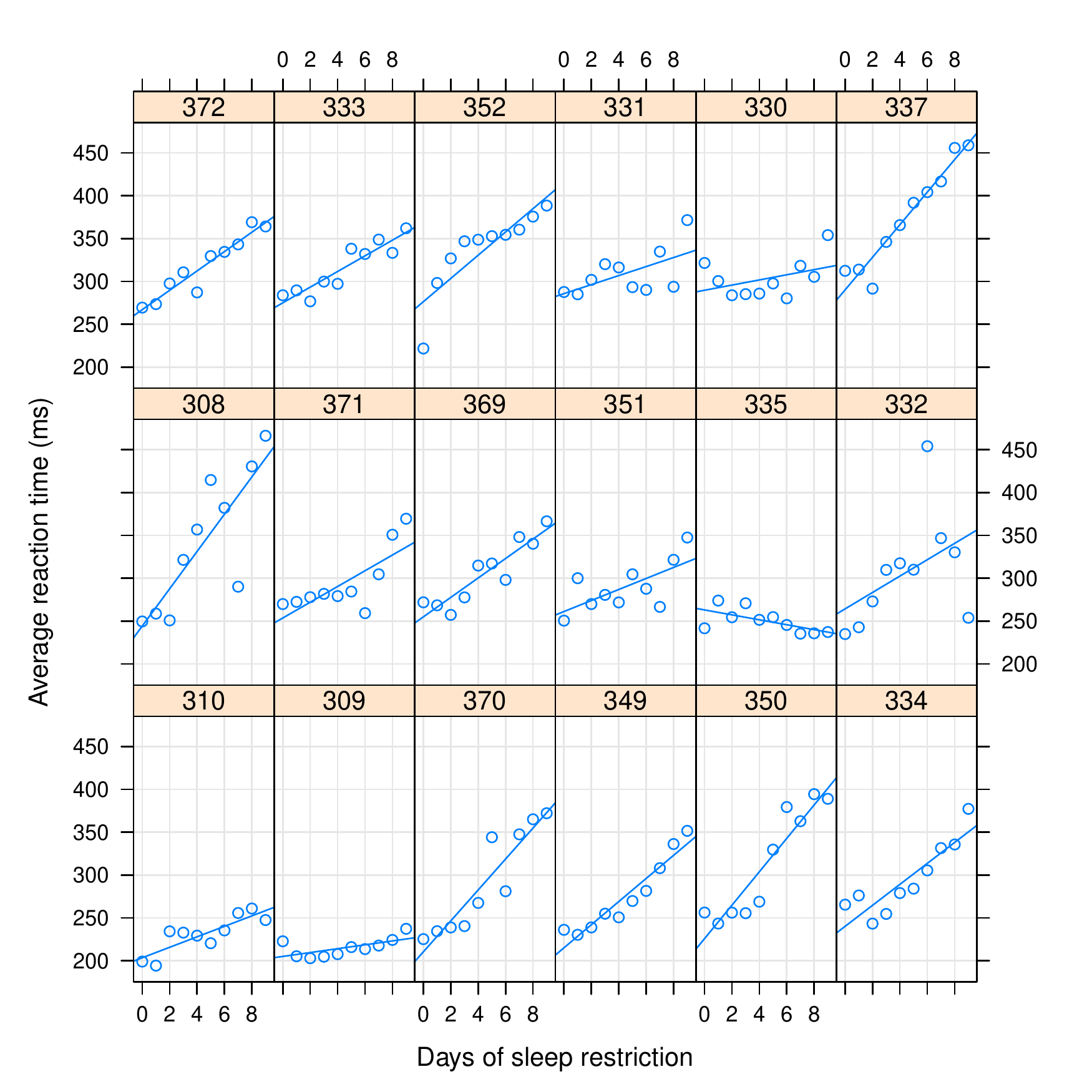} 

}

\caption[Lattice plot of the sleepstudy data]{Lattice plot of the sleepstudy data. For each volunteer there is one panel. The identification number of each volunteer is in the heading of the panels. In the panels the reaction time is plotted against the days of sleep restriction and a regression line is added for each volunteer/panel.}\label{fig:sleepPlot}
\end{figure}
\end{Schunk}

The conditional AIC can be used to find the model that best predicts future 
observations, assuming that future observations share the same random effects as 
the ones used for the model fitting. In case of this data set, using the cAIC 
for model choice corresponds to finding the model that best predicts future 
reaction times of the volunteers that took part in the study.\\ 
After looking at the lattice plot, a first model that could be applied is a model
with a random intercept and a random slope for \code{Days} within each 
volunteer (\code{Subject}): 

\begin{equation}
\label{sleepMod}
  y_{ij} = \beta_0 + \beta_1 \cdot \mbox{day}_{ij} + 
  u_{j0} + u_{j1} \cdot \mbox{day}_{ij} + \epsilon_{ij}
\end{equation}

for $i= 1, \ldots, 18$ and $j = 1, \ldots 10$, with

\[
  \begin{pmatrix}
    u_{j0}\\
    u_{j1}
  \end{pmatrix}
  \sim
  \mathcal{N} \left(
  \begin{pmatrix}
    0\\
    0
  \end{pmatrix}, 
  \begin{pmatrix}
    \tau_1^2 & \tau_{12}^2\\
    \tau_{12}^2 & \tau_2^2
  \end{pmatrix}
  \right).
\]

In the preceding notation $\tau_1^2 = \theta_1$, $\tau_2^2 = \theta_2$ and 
$\tau_{12}^2 = \theta_3$. That $\tau_{12}^2$ is not necessarily zero indicates, 
that the random intercept and the random slope are allowed to be correlated. 

\begin{Schunk}
\begin{Sinput}
R> (m1 <- lmer(Reaction ~ 1 + Days + (1 + Days|Subject), sleepstudy))
\end{Sinput}
\begin{Soutput}
Linear mixed model fit by REML ['lmerMod']
Formula: Reaction ~ 1 + Days + (1 + Days | Subject)
   Data: sleepstudy
REML criterion at convergence: 1743.628
Random effects:
 Groups   Name        Std.Dev. Corr
 Subject  (Intercept) 24.740       
          Days         5.922   0.07
 Residual             25.592       
Number of obs: 180, groups:  Subject, 18
Fixed Effects:
(Intercept)         Days  
     251.41        10.47  
\end{Soutput}
\end{Schunk}

The output shows that the within-subject correlation between the random 
intercepts $u_{j0}$ and the random slopes $u_{j1}$ is low, being estimated as 
$0.07$. Hence there seems to be no evidence that the initial reaction time of 
the volunteers has systematic impact on the pace of increasing reaction time 
following the sleep restriction.\\
Consequently a suitable model might be one in which the correlation structure 
between both is omitted. The model for the response therefore stays 
the same as in (\ref{sleepMod}), but the random effects covariance structure is 
predefined as

\[
  \begin{pmatrix}
    u_{j0}\\
    u_{j1}
  \end{pmatrix}
  \sim
  \mathcal{N} \left(
  \begin{pmatrix}
    0\\
    0
  \end{pmatrix}, 
  \begin{pmatrix}
    \tau_0^2 & 0\\
    0 & \tau_1^2
  \end{pmatrix}
  \right).
\]

Such a model without within-subject correlation is called by

\begin{Schunk}
\begin{Sinput}
R> (m2 <- lmer(Reaction ~ 1 + Days + (1|Subject) + (0 + Days|Subject), 
+ sleepstudy))
\end{Sinput}
\begin{Soutput}
Linear mixed model fit by REML ['lmerMod']
Formula: Reaction ~ 1 + Days + (1 | Subject) + (0 + Days | Subject)
   Data: sleepstudy
REML criterion at convergence: 1743.669
Random effects:
 Groups    Name        Std.Dev.
 Subject   (Intercept) 25.051  
 Subject.1 Days         5.988  
 Residual              25.565  
Number of obs: 180, groups:  Subject, 18
Fixed Effects:
(Intercept)         Days  
     251.41        10.47  
\end{Soutput}
\end{Schunk}

Notice that the estimates of standard deviations of the random effects do not 
differ much between the first and the second model. To decide which model is 
more appropriate in terms of subject specific prediction the conditional AIC can 
be used. Calling the \code{cAIC}-function from the \pkg{cAIC4}-package 
gives the output:

\begin{Schunk}
\begin{Sinput}
R> cAIC(m1)
\end{Sinput}
\begin{Soutput}
$loglikelihood
[1] -824.507

$df
[1] 31.30192

$reducedModel
NULL

$new
[1] FALSE

$caic
[1] 1711.618
\end{Soutput}
\end{Schunk}

The conditional log-likelihood and the corrected degrees of freedom, i.e., the bias correction, are the
first two elements of the resulting list. The third element is called 
\code{reducedModel} and is the model without the random effects covariance 
parameters that were estimated to lie on the boundary of the parameter space, 
see Appendix~\ref{sec:boundary} and \citet{Greven.2010}, and \code{NULL} if there were none on the
boundary. The fourth element says if such a new model was fitted because of the
boundary issue, which was not the case here. The last element is the conditional 
AIC as proposed in \cite{Greven.2010}.\\
The cAIC of the second model \code{m2} is:

\begin{Schunk}
\begin{Sinput}
R> cAIC(m2)$caic
\end{Sinput}
\begin{Soutput}
[1] 1710.426
\end{Soutput}
\end{Schunk}

From a conditional perspective, the second model is thus preferred to the first 
one. This confirms the assertion that the within-subject correlation can be 
omitted in the model.\\
There are several further possible models for these data. For instance the
random slope could be excluded from the model. In this model the pace of 
increasing reaction time does not systematically vary between the volunteers. 
This model is estimated by

\begin{Schunk}
\begin{Sinput}
R> m3 <- lmer(Reaction ~ 1 + Days + (1|Subject), sleepstudy)
\end{Sinput}
\end{Schunk}

The conditional AIC of this model is

\begin{Schunk}
\begin{Sinput}
R> cAIC(m3)$caic
\end{Sinput}
\begin{Soutput}
[1] 1767.118
\end{Soutput}
\end{Schunk}

This is by far larger than the cAIC for the two preceding models. The lattice plot
in Figure~\ref{fig:sleepPlot} already indicated that there is strong evidence
of subject-specific (random) slopes. This is also reflected by the cAIC.\\ 
The conditional AIC is also appropriate for choosing between a simple null model
without any random effects and a complex model incorporating random effects, as 
has been noticed by \cite{Greven.2010}. Thus it is possible to compare the cAIC 
of the three previous mixed models with the standard AIC for a linear model, here 
including three parameters (intercept, linear effect for \code{Days} and error variance)

\begin{Schunk}
\begin{Sinput}
R> -2 * logLik(lm(Reaction ~ 1 + Days, sleepstudy), REML = TRUE)[1] + 2 * 3
\end{Sinput}
\begin{Soutput}
[1] 1899.664
\end{Soutput}
\end{Schunk}

In this case, however, the mixed model structure is evident, reflected by the 
large AIC for the linear model.

\subsection*{Example for generalized linear mixed models} \label{sec:cAIC_posBin}

The \pkg{cAIC4}-package additionally offers a conditional AIC for 
conditionally Poisson distributed responses and an approximate conditional AIC 
for binary data. The Poisson cAIC uses the bias correction (\ref{eq:cAICpo}) and 
the bias correction term for the binary data is (\ref{eq:BerBC}).\\
Making use of the fast \code{refit()} function of the \pkg{lme4}-package, 
both cAICs can be computed moderately fast, since $n - d$ and $n$ model refits 
are required, respectively, with $n$ being the number of observations and $d$ the number of 
responses that are zero for the Poisson responses. In the following, the cAIC 
for Poisson response is computed for the \code{grouseticks} data set from the 
\pkg{lme4}-package as an illustration.\\
The \code{grouseticks} data set was originally published in 
\cite{Elston.2001}. It contains information about the aggregation of parasites, 
so-called sheep ticks, on red grouse chicks. The variables in the data set are 
given in Table~\ref{tab:grouse}. Every chick, identified by \code{INDEX}, is of a 
certain \code{BROOD} and every \code{BROOD}, in turn, corresponds to a specific \code{YEAR}.

\begin{table}[h]
\begin{center}
\begin{tabular}{cc}
\hline
 Variable  & Description \\
\hline
 \code{INDEX}   &  identifier of the chick\\
 \code{TICKS}   &  the number of ticks sampled\\
 \code{BROOD}   &  the brood number\\
 \code{HEIGHT}  &  height above sea level in meters\\
 \code{YEAR}    &  the year as 95, 96 or 97\\
 \code{LOCATION}&  the geographic location code\\
\hline
\end{tabular}
\caption{The variables and response of the grouseticks data set.}
\label{tab:grouse}
\end{center}
\end{table}

The number of ticks is the response variable. Following the authors in a first model the expected 
number of ticks $\lambda_{l}$ with \code{INDEX} $(l)$ is modelled depending on the year and the 
height as fixed effects and for each of the grouping variables \code{BROOD} $(i)$, 
\code{INDEX} $(j)$ and \code{LOCATION} $(k)$ a random intercept is incorporated. 
The full model is

\begin{equation}
\label{eq:Pois}
  \log\left(\mathbb{E}\left(\mbox{\code{TICKS}}_{l}\right)\right) = 
  \log\left(\lambda_{l}\right) = \beta_0 + 
  \beta_1 \cdot \mbox{\code{YEAR}}_{l} + 
  \beta_2 \cdot \mbox{\code{HEIGHT}}_{l} + u_{1,i} + u_{2,j} + u_{3,k}
\end{equation}

with random effects distribution

\[
  \begin{pmatrix}
    u_{1,i}\\
    u_{2,j}\\
    u_{3,k}
  \end{pmatrix}
  \sim
  \mathcal{N} \left(
  \begin{pmatrix}
    0\\
    0\\
    0
  \end{pmatrix}, 
  \begin{pmatrix}
    \tau_1^2 & 0 & 0\\
    0 & \tau_2^2 & 0\\
    0 & 0 & \tau_3^2
  \end{pmatrix}
  \right).
\]

Before fitting the model the covariates \code{HEIGHT} and \code{YEAR} are
centred for numerical reasons and stored in the data set \code{grouseticks\_cen}.

\begin{Schunk}
\begin{Sinput}
R> formula <- TICKS ~ YEAR + HEIGHT + (1|BROOD) + (1|INDEX) + (1|LOCATION)
R> p1  <- glmer(formula, family = "poisson", data = grouseticks_cen)
\end{Sinput}
\end{Schunk}

A summary of the estimated model is given below. Notice that the reported AIC in 
the automated summary of \pkg{lme4} is not appropriate for conditional model 
selection.

%
%
%
%
%

\begin{Soutput}
Generalized linear mixed model fit by maximum likelihood 
(Laplace Approximation) ['glmerMod']
 Family: poisson  ( log )
Formula: TICKS ~ YEAR + HEIGHT + (1 | BROOD) + (1 | INDEX) + (1 | LOCATION)
   Data: grouseticks_cen

     AIC      BIC   logLik deviance df.resid 
  1845.5   1869.5   -916.7   1833.5      397 

Scaled residuals: 
    Min      1Q  Median      3Q     Max 
-1.6507 -0.5609 -0.1348  0.2895  1.8518 

Random effects:
 Groups   Name        Variance  Std.Dev. 
 INDEX    (Intercept) 2.979e-01 5.458e-01
 BROOD    (Intercept) 1.466e+00 1.211e+00
 LOCATION (Intercept) 5.411e-10 2.326e-05
Number of obs: 403, groups:  INDEX, 403; BROOD, 118; LOCATION, 63

Fixed effects:
             Estimate Std. Error z value Pr(>|z|)    
(Intercept)  0.472353   0.134712   3.506 0.000454 ***
YEAR        -0.480261   0.166128  -2.891 0.003841 ** 
HEIGHT      -0.025715   0.003772  -6.817 9.32e-12 ***
---
Signif. codes:  0 *** 0.001 ** 0.01 * 0.05 .

Correlation of Fixed Effects:
       (Intr) YEAR 
YEAR   0.089       
HEIGHT 0.096  0.061 
\end{Soutput}

The conditional log-likelihood and the degrees of freedom for the conditional
AIC with conditionally Poisson distributed responses as in (\ref{eq:cAICpo}) for 
model (\ref{eq:Pois}) are obtained by the call of the \code{cAIC}-function:

\begin{Schunk}
\begin{Sinput}
R> set.seed(42)
R> cAIC(p1)
\end{Sinput}
\end{Schunk}

\begin{Soutput}
$loglikelihood
[1] -572.0133

$df
[1] 205.5786

$reducedModel
NULL

$new
[1] FALSE

$caic
[1] 1555.184
\end{Soutput}

The output is the same as for Gaussian linear mixed models. It becomes apparent 
that there is a substantial difference between the conditional and the marginal 
AIC: In the output of the model the marginal AIC is reported to be 1845.48. 
Note that the marginal AIC is biased, see \cite{Greven.2010}, 
and based on  a different likelihood
.\\ 
In the full model, the standard deviations of the random effects are
rather low. It thus may be possible to exclude one of the grouping variables
from the model, only maintaining two random effects. There are three possible
models with one of the random effects terms excluded.\\ 
If the random intercept associated with \code{LOCATION} is excluded the model 
is

\begin{Schunk}
\begin{Sinput}
R> formel <- TICKS ~ YEAR + HEIGHT + (1|BROOD) + (1|INDEX)
R> p2  <- glmer(formel, family = "poisson", data = grouseticks_cen)
R> cAIC(p2)$caic
\end{Sinput}
\end{Schunk}

\begin{Soutput}
[1] 1555.214
\end{Soutput}

The conditional AIC is almost the same as for the full model. It may thus make 
sense to choose the reduced model and for the prediction of the number of ticks 
not to make use of the random intercept associated with the \code{LOCATION} 
grouping.\\
Another possible model can be obtained by omitting the random intercepts for the 
\code{INDEX} grouping structure instead of those associated with 
\code{LOCATION}. This would make the model considerably simpler, since each 
chick has an \code{INDEX} and hence a random intercept is estimated for each 
observation in order to deal with overdispersion in the data.

\begin{Schunk}
\begin{Sinput}
R> formel <- TICKS ~ YEAR + HEIGHT + (1|BROOD) + (1|LOCATION)
R> p3  <- glmer(formel, family = "poisson", data = grouseticks_cen)
R> cAIC(p3)$caic
\end{Sinput}
\end{Schunk}

\begin{Soutput}
[1] 1842.205
\end{Soutput}

The large cAIC in comparison with the two preceding models documents that the
subject-specific random intercept for each observation should be included.\\
The final model for the comparison omits random intercepts associated with the 
\code{BROOD} grouping. This is equivalent to setting the associated random 
intercepts variance to zero, i.e., $\tau_2^2 = 0$. 

\begin{Schunk}
\begin{Sinput}
R> formel <- TICKS ~ YEAR + HEIGHT + (1|INDEX) + (1|LOCATION)
R> p4  <- glmer(formel, family = "poisson", data = grouseticks_cen)
R> cAIC(p4)$caic
\end{Sinput}
\end{Schunk}

\begin{Soutput}
[1] 1594.424
\end{Soutput}

The cAIC is higher than the cAICs for the full model and the model without the 
\code{LOCATION} grouping structure. Consequently either the full model or 
the model without the \code{LOCATION} grouping structure is favoured by the 
cAIC. The authors favour the latter.

\section{A scheme for stepwise conditional variable selection}

Now having the possibility to compare different (generalized) linear mixed models 
via the conditional AIC, we introduce a model selection procedure in 
this section, searching the space of possible model candidates in a stepwise 
manner. Inspired by commonly used \code{step}-functions as for example given 
by the \code{stepAIC} function in the \pkg{MASS}-package \citep{MASS}, our 
\code{stepcAIC}-function provides an automatic model selection applicable to 
all models of the class \code{merMod} (produced by 
\code{[g]lmer}) 
or 
objects resulting from a \pkg{gamm4}-call. 

For example, consider the sleepstudy model

\begin{Schunk}
\begin{Sinput}
R> fm1 <- lmer(Reaction ~ Days + (Days | Subject), sleepstudy)
\end{Sinput}
\end{Schunk}

\noindent which implicitly fits the random effects structure 
\code{(1 + Days | Subject)} (correlated random intercept and slope). 
In order to perform a data-driven search for the 
best model, a backward step procedure needs to fit and evaluate the following 
three nested models (uncorrelated random intercept and slope, only random slope, only random intercept).

\begin{Schunk}
\begin{Sinput}
R> fm1a <- lmer(Reaction ~ Days + (1 | Subject) + (0 + Days | Subject), 
+   sleepstudy)
R> fm1b <- lmer(Reaction ~ Days + (0 + Days | Subject), sleepstudy)
R> fm1c <- lmer(Reaction ~ Days + (1 | Subject), sleepstudy)
\end{Sinput}
\end{Schunk}

Choosing the model \code{fm1a} in the first step, further model comparisons 
may be performed by for example reducing the model once again or adding another 
random effect. For this purpose, the \code{stepcAIC}-function provides the 
argument \code{direction}, having the options \code{backward}, 
\code{forward} and \code{both}. Whereas the \code{backward}- and 
\code{forward}-direction procedures fit and evaluate all nested or extended 
models step-by-step, the \code{both}-direction procedure alternates between 
forward- and backward-steps as long as any of both steps lead to an improvement 
in the cAIC. During model modifications in each step, the function allows to 
search through different types of model classes.\\
For fixed effects selection, the step procedure furthermore can be used to 
successively extend or reduce the model in order to check whether a fixed 
effect has a constant, linear or non-linear impact. For example, we specify a 
generalized additive mixed model (GAMM) as follows \citep[cf.][]{Gu.1991}

$$y_{ij} = \beta_0 + x_{1,i,j} \beta_1 + f(x_{3,i,j}) + b_i + \varepsilon_{ij}, \quad i=1,\ldots,20, j=1,\ldots,J_i,$$

with metric variables $x_1$ and $x_3$ 
in the \code{guWahbaData} supplied in the \pkg{cAIC4} package with continuous covariates 
$x_0, x_1, x_2$ and $x_3$.
.

The corresponding model fit in \textsf{R} using \pkg{gamm4} is given by

\begin{Schunk}
\begin{Sinput}
R> set.seed(42)
R> guWahbaData$fac <- fac <- as.factor(sample(1:20, 400, replace =TRUE))
R> guWahbaData$y <- guWahbaData$y + model.matrix(~ fac - 1) %*% rnorm(20) * 0.5
R> br <- gamm4(y ~ x1 + s(x3, bs = "ps"), data = guWahbaData, random = ~ (1|fac))
\end{Sinput}
\end{Schunk}

\noindent resulting in the following non-linear estimate of $f(x_{3,i,j})$ (Figure~\ref{fig:guWahbasx}).

\begin{Schunk}
\begin{figure}

{\centering \includegraphics[width=\maxwidth]{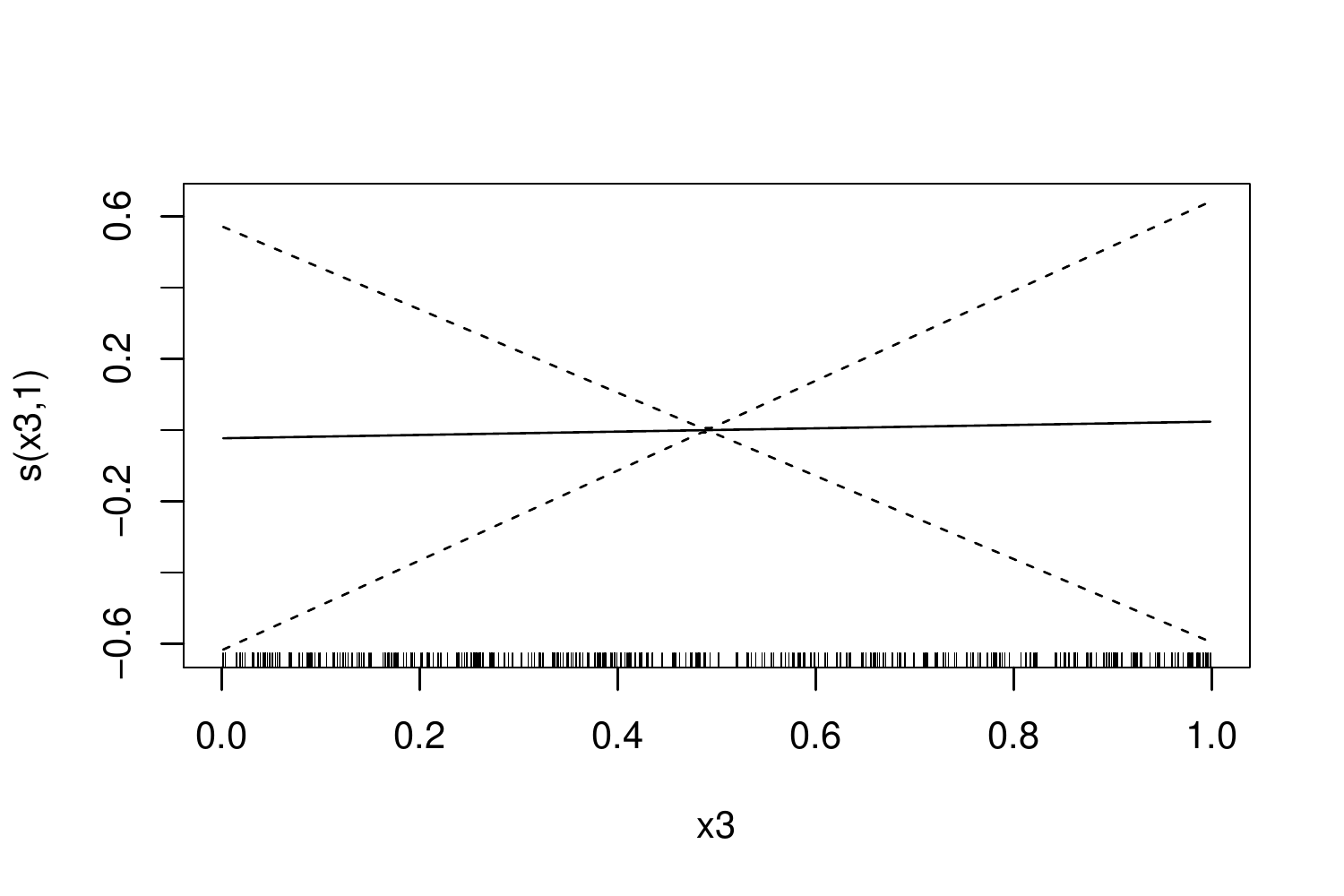} 

}

\caption[Plot of non-linear effect estimate for covariate $x_3$]{Plot of non-linear effect estimate for covariate $x_3$.}\label{fig:guWahbasx}
\end{figure}
\end{Schunk}

Applying the backward stepwise procedure to the model \code{br} via

\begin{Schunk}
\begin{Sinput}
R> stepcAIC(br, trace = TRUE, direction = "backward", data = guWahbaData)
\end{Sinput}
\end{Schunk}

the procedure stops after one step with a warning, saying that the model contains zero variance components and the corresponding terms must be removed manually. This is due to the fact that the \code{stepcAIC} function can not reduce non-linear effects such as $f(x_{3,i,j})$ automatically, as the type of additive effect depends on the specification of the \code{s}-term and its arguments. Modifying the term manually, a GLMM is fitted and passed to the \code{stepcAIC} function.

\begin{Schunk}
\begin{Sinput}
R> br0 <- gamm4(y ~ x1 + x3, data = guWahbaData, random = ~ (1|fac))
R> stepcAIC(br0, trace = TRUE, direction = "backward", data = guWahbaData)
\end{Sinput}
\end{Schunk}

In the next steps \code{stepcAIC} removes \code{x3} completely from the model and 
also checks whether a GLM with no random effects at all might be the best possible model, 
hence having searched for the smallest cAIC in three different model classes in the end.\\

\color{black}
Whereas the backward procedure has straightforward mechanism and does not
need any further mandatory arguments as shown in the previous example, the
\code{stepcAIC}-function provides several optional and obligatory arguments
for the \code{forward}- and \code{both}
procedure in order to limit the possibly large number of model extensions.
Regarding the required parameters, the user must specify the variables, which
may be added with fixed or random effects as they are referred to in the
\code{data.frame} given by the argument \code{data}. For the fixed effects, 
this is done by specifying the \code{fixEf} argument, which expects a 
character vector with the names of the covariates, e.g., 
\code{fixEf=c("x1","x2")}. Variables listed in the \code{fixEf}-argument are
firstly included in the model as linear terms and, if the linear effect leads to
an improvement of the cAIC, checked for their non-linearity by evaluating the
cAIC of the corresponding model(s). Model extensions
resulting from additional random effects are created in two different ways. A
new model may, on the one hand, include a random intercept for a variable forming a grouping structure
(in the \code{sleepstudy} example for \code{Subject}) or, on the other hand, a random slope for a variable (\code{Days} in this case). These two types are
specified using the arguments \code{groupCandidates} for grouping variables candidates or
\code{slopeCandidates} for candidates for variables with random slope, again by referring to the variable names in
\code{data} as string.

Further optional arguments can determine the way random effects are treated in 
the step procedure:

\begin{itemize}

\item[-] \code{allowUseAcross}: logical value whether slope variables, which are already in use with a grouping variable can also be used with other grouping variables,   

\item[-] \code{maxSlopes}: maximum number of slopes for one 
grouping variable.


\end{itemize}

Following the \code{stepAIC}-function, the \code{stepcAIC}-function also
provides an argument for printing interim results (\code{trace}) and allows for the remaining terms of the initial model
to be unaffected by the procedure (\code{keep}: list with entries
\code{fixed} and \code{random}, each either \code{NULL} or a
\code{formula}). In addition, the user may choose whether the cAIC is
calculated for models, for which the fitting procedure in
(\code{g})\code{lmer} could not find an optimum (\code{calcNonOptimMod, default = FALSE})
and might choose the type of smoothing terms added in forward steps
(\code{bsType}).

If the step-function is used for large datasets or in the presence of
highly complex models the fitting procedures as well as the calculations of the
cAIC can be parallelized by defining the number of cores (\code{numCores})
being used if more than one model has to be fitted and evaluated in any step
(therefore passing the \code{numCores}-argument to a
\code{mclapply}-function implemented in the \code{parallel}-package
\citep{R.2016}).\\

Due to the variety of additive model definitions in \pkg{gamm4}, the
\code{stepcAIC} is however limited in its generic step-functionality for
GAMMs. On the one hand, extensions with non-linear effects are restricted to one
smooth class given by \code{bsType}, on the other hand, the step-procedure is
not able to deal with further arguments passed in smooth terms. The latter point
is a current limitation, since the default basis dimension of the
smooth term (i.e., the number of knots and the order of the penalty) is
essentially arbitrary. 

An additional current limitation of the \code{stepcAIC}-function in its 
applications with GAMMs is the handling of zero variance components occurring during the 
function call. As a meaningful handling of 
zero variance smoothing terms would depend on the exact specification of the 
non-linear term, the stepwise procedure is stopped and returns the 
result of the previous step. After removing the zero variance term manually the 
user may call the step-function again.

\subsection*{Examples}

In order to demonstrate some functionalities of the \code{stepcAIC}-function,
various examples are given in the following using the \code{Pastes} data set
\citep{Davies.1972}, which is available in the \pkg{lme4}-package. The data 
set consists of 60 observations including one metric variable \code{strength}, 
which is the strength of a chemical paste product and the categorical 
variables \code{batch} (the delivery batch), the \code{cask} within the delivery batch and \code{sample}, which is an identifier from what cask in what batch the paste sample was taken. 

Starting with a random effects backward selection, the model \code{fm3}

\begin{Schunk}
\begin{Sinput}
R> fm3 <- lmer(strength ~ 1 + (1|sample) + (1|batch), Pastes)
\end{Sinput}
\end{Schunk}

may be automatically reduced using

\begin{Schunk}
\begin{Sinput}
R> fm3_step <- stepcAIC(fm3, direction = "backward", trace = TRUE, data = Pastes)
\end{Sinput}
\end{Schunk}

\begin{Soutput}
Starting stepwise procedure...
_____________________________________________
_____________________________________________

Step 1 (backward):  cAIC=178.2809
Best model so far: ~ (1 | sample) + (1 | batch)
New Candidates:

Calculating cAIC for 2 model(s) ...

        models loglikelihood        df     caic
  ~(1 | batch)    -141.49709  9.157892 301.3100
 ~(1 | sample)     -58.95458 30.144477 178.1981

_____________________________________________
_____________________________________________

Step 2 (backward):  cAIC=178.1981
Best model so far: ~ (1 | sample)
New Candidates:

Calculating cAIC for 1 model(s) ...

 models loglikelihood df     caic
     ~1     -155.1363  2 312.2727

_____________________________________________
_____________________________________________

Best model:  ~ (1 | sample) , cAIC: 178.1981 
_____________________________________________
\end{Soutput}

where in a first step, the random intercept of \code{batch} is dropped.
Afterwards, the procedure compares the cAICs of the models \code{lmer(strength
~ 1 + (1|sample), Pastes)} and \code{lm(strength ~ 1, Pastes)},
keeping the second random effect due to a smaller cAIC of the linear mixed
model.

Using the step function the other way round, a forward stepwise selection can be
 initialized by a simple linear model

\begin{Schunk}
\begin{Sinput}
R> fm3_min <- lm(strength ~ 1, data = Pastes)
\end{Sinput}
\end{Schunk}

followed by a \code{stepcAIC}-call

\begin{Schunk}
\begin{Sinput}
R> fm3_min_step <- stepcAIC(fm3_min, 
+   groupCandidates = c("batch", "sample"),
+   direction = "forward", trace = TRUE, 
+   data = Pastes, analytic = TRUE)
\end{Sinput}
\end{Schunk}

where possible new candidates for grouping variables are specified using the
\code{groupCandidates}-argument. Again, the random intercept model with
group \code{sample} is finally selected. 

To illustrate the use of the \code{stepcAIC}-function in the context of GAMM
selection, two examples are generated following the \pkg{gamm4}-help page on the basis of the \code{guWahbaData} data set. First, the GAMM $$y_{ij} = \beta_0 + f(x_{0,i,j}) + x_{1,i,j} \beta_1 + f(x_{2,i,j}) + b_i, \quad i=1,\ldots,20, j=1,\ldots,J_i$$ is fitted to the \code{guWahbaData} including a nonlinear term for the covariate \code{x0} using a thin-plate regression spline, a P-spline \citep{Eilers.1996} for the covariate \code{x2} as well as a random effect for the grouping variable \code{fac}.

\begin{Schunk}
\begin{Sinput}
R> br <- gamm4(y ~ s(x0) + x1 + s(x2, bs = "ps"), 
+   data = guWahbaData, random = ~ (1|fac))
\end{Sinput}
\end{Schunk}

In order to check for linear or non-linear effects of the two other covariates 
\code{x1} and \code{x3}, the \code{stepcAIC}-function is employed.

\begin{Schunk}
\begin{Sinput}
R> br_step <- stepcAIC(br, fixEf = c("x1", "x3"), 
+   direction = "both", 
+   data = guWahbaData)
\end{Sinput}
\end{Schunk}

After changing the linear effect \code{x1} to a non-linear
effect, i.e., \code{s(x1, bs = "tp")}, and therefore improving the model's
cAIC in a first forward step, the function stops due to zero variance components.

The final model \code{br\_step} to this point is thus given by \code{y
~ s(x0, bs = "tp") + s(x2, bs = "ps") + s(x1, bs = "tp") + (1 | fac)}.
In contrast to the effect of covariate \code{x2} modeled as P-spline, the effects of 
covariates \code{x0} and \code{x1} are modeled as thin plate regression
splines \citep{Wood.2017}. For \code{x0}, this is due to the initial model definition, as
\code{s(x0)} is internally equal to \code{s(x0, bs = "tp")}, whereas for \code{x1}, 
the definition of the spline is set by the argument \code{bsType}
of the \code{stepcAIC}-function. As the \code{bsType}-argument is not
specified in the call, the default \code{"tp"} is used.

Finally, a demonstration of the \code{keep}-statement is given for the model

\begin{Schunk}
\begin{Sinput}
R> br2 <- gamm4(y ~ s(x0, bs = "ps") + x2, data = guWahbaData, 
+   random = ~ (1|fac))
\end{Sinput}
\end{Schunk}
 
where the aim is to prevent the step procedure changing the linear effect of the
covariate \code{x2}, the non-linear effect of \code{x0} as well as the 
random effect given by \code{~ (1|fac)}.

\begin{Schunk}
\begin{Sinput}
R> br2_step <- stepcAIC(br2, trace = TRUE, direction = "both",
+   fixEf = c("x1", "x3"), bsType = "cs",
+   keep = list(fixed = ~ s(x0, bs = "ps") + x2,
+   random= ~ (1|fac)), data = guWahbaData)
\end{Sinput}
\end{Schunk}

After successively adding a linear effect of \code{x1} to the model, neither the following backward step nor
another forward step do improve the cAIC. The final model is given by 
\code{y ~ s(x0, bs = "ps") + x1 + x2} and random effect \code{(1|fac)}.

\section{Conclusion}
This paper gives a hands-on introduction to the \textsf{R}-package \pkg{cAIC4} allowing for 
model selection in mixed models based on the conditional AIC. The
package and the paper offer a possibility for users from the empirical sciences to use 
the conditional AIC without having to worry about lengthy and complex 
calculations or mathematically sophisticated boundary issues of the parameter 
space. The applications presented in this paper go far beyond model selection 
for mixed models and extend to penalized spline smoothing and other structured 
additive regression models. Furthermore a stepwise algorithm for these models is 
introduced that allows for fast model selection.\\
Often statistical modelling is not about finding one 'true model'. In such cases it is of interest 
to define weighted sums of plausible models. This 
approach called model averaging is presented in \cite{Zhang.2014} for weights chosen by the 
cAIC. We plan to implement this approach in \pkg{cAIC4}. Another future research path 
is to implement an appropriate version of the Bayesian information criterion 
(BIC) for conditional model selection.\\

\textbf{Acknowledgements}\\
The research by Thomas Kneib and Benajmin S\"afken was supported by the RTG
1644 - Scaling Problems in Statistics and the Centre for Statistics at Georg-August-Universit\"at G\"ottingen.
Sonja Greven and David R\"ugamer acknowledge funding by Emmy Noether grant GR 3793/1-1 
from the German Research Foundation.

\appendix

\section{Dealing with the boundary issues}
\label{sec:boundary}

A major issue in obtaining the conditional AIC in linear mixed models is to account for potential 
parameters of $\boldsymbol{\theta}$ on the boundary of the parameter space \citep[see][]{Greven.2010}. This 
needs to be done in order to ensure positive definiteness of the covariance 
matrix $\boldsymbol{D}_{\boldsymbol{\theta}}$.\\ 
The restructuring of the model in order to obtain the cAIC is done automatically 
by \pkg{cAIC4}. To gain insight into the restructuring, an understanding of 
the mixed model formulas used in \pkg{lme4} is essential. For an in depth 
explanation on how the formula module of \pkg{lme4} works, see 
\cite{lme4.2015}, Section 2.1.\\
Suppose we want to fit a mixed model with two grouping factors \code{g1} and 
\code{g2}. Within the first grouping factor \code{g1}, there are three 
continuous variables \code{v1}, \code{v2} and \code{v3} and within the 
second grouping factor there is only one variable \code{x}. Thus there are not 
only random intercepts but also random slopes that are possibly correlated 
within the groups. Such a model with response \code{y} would be called in 
\pkg{lme4} by

\begin{Schunk}
\begin{Sinput}
R> m <- lmer(y ~ (v1 + v2 + v3|g1) + (x|g2), exampledata)
\end{Sinput}
\end{Schunk}

In mixed models fitted with \pkg{lme4}, the random effects covariance 
matrix $\boldsymbol{D}_{\boldsymbol{\theta}}$ always has block-diagonal 
structure. For instance in the example from above the Cholesky factorized blocks 
of 
the estimated 
$\boldsymbol{D}_{\boldsymbol{\theta}}$ associated with each random effects 
term are 

\begin{Schunk}
\begin{Sinput}
R> getME(m, "ST")
\end{Sinput}
\begin{Soutput}
$g2
            [,1] [,2]
[1,]  1.18830353  NaN
[2,] -0.01488359    0

$g1
              [,1]        [,2] [,3] [,4]
[1,]  1.0184626697  0.00000000  NaN  NaN
[2,] -0.1438761295  0.05495809  NaN  NaN
[3,] -0.0007341796  0.19904339    0  NaN
[4,] -0.0883652598 -1.36463267 -Inf    0
\end{Soutput}
\end{Schunk}

If any of the diagonal elements of the blocks are zero the corresponding random 
effects terms are deleted from the formula. In \pkg{lme4} this is done 
conveniently by the component names list 

\begin{Schunk}
\begin{Sinput}
R> m@cnms
\end{Sinput}
\begin{Soutput}
$g2
[1] "(Intercept)" "x"          

$g1
[1] "(Intercept)" "v1"          "v2"          "v3"         
\end{Soutput}
\end{Schunk}

Thus a new model formula can be obtained by designing a new components names 
list:

\begin{Schunk}
\begin{Sinput}
R> varBlockMatrices <- getME(m, "ST")
R> cnms <- m@cnms
R> for(i in 1:length(varBlockMatrices)){
+   cnms[[i]] <- cnms[[i]][which(diag(varBlockMatrices[[i]]) != 0)]
+ }
R> cnms
\end{Sinput}
\begin{Soutput}
$g2
[1] "(Intercept)"

$g1
[1] "(Intercept)" "v1"         
\end{Soutput}
\end{Schunk}

The \code{cnms2formula} function from the \pkg{cAIC4}-package forms a new 
formula from the \code{cnms} object above. Hence the new formula can be 
computed by

\begin{Schunk}
\begin{Sinput}
R> rhs  <- cAIC4:::cnms2formula(cnms)
R> lhs  <- formula(m)[[2]]
R> reformulate(rhs, lhs)
\end{Sinput}
\begin{Soutput}
y ~ (1 | g2) + (1 + v1 | g1)
\end{Soutput}
\end{Schunk}

This code is called from the \code{deleteZeroComponents} function in the 
\pkg{cAIC4}-package. This function automatically deletes all zero components 
from the model. The\\
\code{deleteZeroComponents} function is called recursively, so the new model 
is checked again for zero components. In the example above only the random 
intercepts are non-zero. Hence the formula of the reduced model from which the 
conditional AIC is calculated is

\begin{Schunk}
\begin{Sinput}
R> formula(cAIC4:::deleteZeroComponents(m))
\end{Sinput}
\begin{Soutput}
y ~ (1 | g2) + (1 | g1)
\end{Soutput}
\end{Schunk}

With the new model the conditional AIC is computed. If there are no random 
effect terms left in the formula, a linear model and the conventional AIC is 
returned. The \code{deleteZeroComponents} function additionally accounts for 
several special cases that may occur.\\
Notice however that in case of using smoothing terms from \pkg{gamm4} no 
automated check for boundary issues can be applied and zero components have 
to be manually deleted.

\section{Computational matters} \label{sec:compMat}

\subsection*{Gaussian responses}

The corrected conditional AIC proposed in \cite{Greven.2010} accounts for the 
uncertainty induced by the estimation of the random effects covariance 
parameters $\boldsymbol{\theta}$. In order to adapt the findings of 
\cite{Greven.2010}, a number of quantities from the \code{lmer} model fit need 
to be extracted and transformed. In the following these computations are 
presented. They are designed to minimize the computational burden and maximize 
the numerical stability. Parts of the calculations needed, for instance the 
Hessian of the ML/REML criterion, can also be found in \cite{lme4.2015}. Notice 
however, that \pkg{lme4} does not explicitly calculate these quantities but 
uses derivative free optimizers for the profile likelihoods.\\
A core ingredient of mixed models is the covariance matrix of the marginal 
responses $\boldsymbol{y}$. The inverse of the scaled covariance matrix 
$\boldsymbol{V}_0$ will be used in the following calculations:  

\[
  \bm{V} = \mbox{cov}(\bm{y}) = \sigma^2 \left( \bm{I}_n +  
  \bm{Z}\bm{\Lambda}_{\bm{\theta}}\bm{\Lambda}_{\bm{\theta}}^t\bm{Z}^t\right) = 
  \sigma^2 \bm{V}_0.
\]

Large parts of the computational methods in \pkg{lme4} rely on a sparse
Cholesky factor that satisfies

\begin{equation}
\label{eq:Chol}
  \boldsymbol{L}_{\boldsymbol{\theta}}\boldsymbol{L}_{\boldsymbol{\theta}}^t =
  \boldsymbol{\Lambda}_{\boldsymbol{\theta}}^t\boldsymbol{Z}^t\boldsymbol{Z}\boldsymbol{\Lambda}_{\boldsymbol{\theta}} + 
  \boldsymbol{I}_q.
\end{equation}

From this equation and keeping in mind that $\boldsymbol{I} - \boldsymbol{V}_0^{-1} = \boldsymbol{Z} 
\left( \boldsymbol{Z}^t \boldsymbol{Z} + \left(\boldsymbol{\Lambda}_{\boldsymbol{\theta}}^{t}\right)^{-1}\boldsymbol{\Lambda}_{\boldsymbol{\theta}}^{-1} \right)^{-1}\boldsymbol{Z}^t$, see 
\cite{Greven.2010}, it follows that

\begin{align*}
  \boldsymbol{\Lambda}_{\boldsymbol{\theta}} \left(\boldsymbol{L}_{\boldsymbol{\theta}}^t\right)^{-1}
  \boldsymbol{L}_{\boldsymbol{\theta}}^{-1} \boldsymbol{\Lambda}_{\boldsymbol{\theta}}^t &=
  \left( \boldsymbol{Z}^t \boldsymbol{Z} + \left(\boldsymbol{\Lambda}_{\boldsymbol{\theta}}^{t}\right)^{-1} \boldsymbol{\Lambda}_{\boldsymbol{\theta}}^{-1}  \right)^{-1}\\
  \Rightarrow \qquad
  \boldsymbol{I} - \boldsymbol{V}_0^{-1}
  &= 
  \left(\boldsymbol{L}_{\boldsymbol{\theta}}^{-1}\boldsymbol{\Lambda}_{\boldsymbol{\theta}}^t \boldsymbol{Z}^t \right)^t
  \left(\boldsymbol{L}_{\boldsymbol{\theta}}^{-1}\boldsymbol{\Lambda}_{\boldsymbol{\theta}}^t \boldsymbol{Z}^t \right).
\end{align*}

Hence the inverse of the scaled variance matrix $\bm{V}_0^{-1}$ can be 
efficiently computed with the help of the \textsf{R}-package \pkg{Matrix} \citep[see][]{Matrix.2017}
that provides methods specifically for sparse matrices: 

\begin{Schunk}
\begin{Sinput}
R> Lambdat <- getME(m, "Lambdat")
R> V0inv <- diag(rep(1, n)) - 
+   crossprod(solve(getME(m, "L"), system = "L") %*% 
+   solve(getME(m, "L"), Lambdat, system = "P") %*% t(Z))
\end{Sinput}
\end{Schunk}

Notice that \code{solve(getME(m, "L"), Lambdat, system = "P")} accounts for a 
fill-reducing permutation matrix $\boldsymbol{P}$ associated (and stored) 
with $\boldsymbol{L}_{\boldsymbol{\theta}}$, see \cite{lme4.2015}, and is thus 
equivalent to

\begin{Schunk}
\begin{Sinput}
R> P %*% Lambdat
\end{Sinput}
\end{Schunk}

Another quantity needed for the calculation of the corrected degrees of freedom 
in the conditional AIC are the derivatives of the scaled covariance matrix of 
the responses $\boldsymbol{V}_0$ with respect to the $j$-th element of the 
parameter vector $\boldsymbol{\theta}$:

\[
  \boldsymbol{W}_j = \frac{\partial}{\partial \theta_j}\boldsymbol{V}_0 =
  \boldsymbol{Z} \boldsymbol{D}_{\boldsymbol{\theta}}^{(j)} \boldsymbol{Z}^t,
\]

where the derivative of the scaled covariance matrix of the random effects with
respect to the $j$-th variance parameter is defined by

\[
  \boldsymbol{D}_{\boldsymbol{\theta}}^{(j)} =  
  \frac{1}{\sigma^2}\frac{\partial}{\partial \theta_j}\boldsymbol{D}_{\boldsymbol{\theta}}.
\]
  
Notice that $\boldsymbol{D}_{\boldsymbol{\theta}} = \left[d_{st}\right]_{s, t= 
1, \ldots,q}$ is symmetric and block-diagonal and its scaled elements are stored 
in $\boldsymbol{\theta}$, hence $d_{st} = d_{ts} = \theta_j \sigma^2$, for 
certain $t, s$ and $j$. Thus the matrix 
$\boldsymbol{D}_{\boldsymbol{\theta}}^{(j)} = \left[ d_{st}^{(j)} \right]_{s,t = 
1, \ldots,q}$ is sparse with

\[
  d_{st}^{(j)} = \begin{cases} 
  				1 &, \mbox{ if } d_{st} = d_{ts} = \theta_j \sigma^2\\
					0 &, \mbox{ else}.
				\end{cases}	
\]

The derivative matrices $\boldsymbol{W}_j$ can be derived as follows:
  
\begin{Schunk}
\begin{Sinput}
R> Lambda <- getME(m, "Lambda")
R> ind    <- getME(m, "Lind")
R> len    <- rep(0, length(Lambda@x))
R>   
R> for(j in 1:length(theta)) {
+   LambdaS                    <- Lambda
+   LambdaSt                   <- Lambdat
+   LambdaS@x                  <- LambdaSt@x                    <- len
+   LambdaS@x[which(ind == j)] <- LambdaSt@x[which(ind == j)]   <- 1
+   diagonal                   <- diag(LambdaS)
+   diag(LambdaS)              <- diag(LambdaSt)                <- 0
+   Dj                         <- LambdaS + LambdaSt
+   diag(Dj)                   <- diagonal
+   Wlist[[j]]                 <- Z %*% Dj %*% t(Z)
+ }
\end{Sinput}
\end{Schunk}


The following matrix is essential to derive the 
corrected AIC of Theorem 3 in \cite{Greven.2010}. Adapting their notation, the 
matrix is

\[
  \boldsymbol{A} =  \boldsymbol{V}_0^{-1} - \boldsymbol{V}_0^{-1}\boldsymbol{X}
  \left( \boldsymbol{X}^t \boldsymbol{V}_0^{-1} \boldsymbol{X} \right)^{-1} \boldsymbol{X}^t \boldsymbol{V}_0^{-1}.
\]

Considering that the cross-product of the fixed effects Cholesky factor is

\[
  \boldsymbol{X}^t \boldsymbol{V}_0^{-1} \boldsymbol{X} = \boldsymbol{R}_{\boldsymbol{X}}^t\boldsymbol{R}_{\boldsymbol{X}},
\]

the matrix $\boldsymbol{A}$ can be rewritten

\[
  \boldsymbol{A} = \boldsymbol{V}_0^{-1} - \left(\boldsymbol{X}\boldsymbol{R}_{\boldsymbol{X}}^{-1}\boldsymbol{V}_0^{-1}\right)
                         \left(\boldsymbol{X}\boldsymbol{R}_{\boldsymbol{X}}^{-1}\boldsymbol{V}_0^{-1}\right)^t.
\]

Accordingly the computation in \textsf{R} can be done as follows:

\begin{Schunk}
\begin{Sinput}
R> A <- V0inv - crossprod(crossprod(X %*% solve(getME(m, "RX")), V0inv))
\end{Sinput}
\end{Schunk}

With these components, the Hessian matrix 

\[
  \boldsymbol{B} = \frac{\partial^2 \mathrm{REML}(\boldsymbol{\theta})}{\partial \boldsymbol{\theta}\partial
  \boldsymbol{\theta}^t} \mbox{ or } \boldsymbol{B} = \frac{\partial^2 \mathrm{ML}(\boldsymbol{\theta})}{\partial
  \boldsymbol{\theta}\partial \boldsymbol{\theta}^t}
\] 

and the matrix

\[
  \boldsymbol{G} = \frac{\partial^2 \mathrm{REML}(\boldsymbol{\theta})}{\partial \boldsymbol{\theta}\partial
  \boldsymbol{y}^t} \mbox{ or } \boldsymbol{G} = \frac{\partial^2 \mathrm{ML}(\boldsymbol{\theta})}{\partial
  \boldsymbol{\theta}\partial \boldsymbol{y}^t},
\]

depending on whether the restricted or the marginal profile log-likelihood $\mathrm{REML}(\bm{\theta})$ or $\mathrm{ML}(\bm{\theta})$ is
used, can be computed straightforward as in \cite{Greven.2010}. Depending on the 
optimization, it may not even be necessary to compute the matrix 
$\boldsymbol{B}$. Considering that $\boldsymbol{B}$ is the Hessian of the profile 
(restricted) log-likelihood, the matrix can also be taken from the model fit, although this is only a numerical approximation. If the Hessian is computed it
is stored in: 

\begin{Schunk}
\begin{Sinput}
R> B <- m@optinfo$derivs$Hessian 
\end{Sinput}
\end{Schunk}

The inverse of $\boldsymbol{B}$ does not need to be calculated -- instead, if
$\boldsymbol{B}$ is positive definite, a Cholesky decomposition and two backward 
solves are sufficient:

\begin{Schunk}
\begin{Sinput}
R> Rchol   <- chol(B)
R> L1      <- backsolve(Rchol, G, transpose = TRUE)
R> Gammay  <- backsolve(Rchol, L1)
\end{Sinput}
\end{Schunk}

The trace of the hat matrix, the first part of the effective degrees of freedom 
needed for the cAIC, can also easily be computed with the help of the residual 
matrix $\boldsymbol{A}$

\begin{Schunk}
\begin{Sinput}
R> df <- n - sum(diag(A))
\end{Sinput}
\end{Schunk}

The correction needed to account for the uncertainty induced by the estimation 
of the variance parameters can be added for each random effects variance 
parameter separately by calculating

\begin{Schunk}
\begin{Sinput}
R> for (j in 1:length(theta)) {
+   df <- df + sum(Gammay[j,] %*% A %*% Wlist[[j]] %*% A %*% y)  
+ }
\end{Sinput}
\end{Schunk}

\subsection*{Poisson responses}

The computation of the bias correction for Poisson distributed responses is 
obtained differently. In a first step the non-zero responses need to be 
identified and a matrix with the responses in each column is created. Consider 
the \code{grouseticks} example in Section~\ref{sec:cAIC_posBin} with
the model \code{p1} fitted by \code{glmer}. 

\begin{Schunk}
\begin{Sinput}
R> y              <- p1@resp$y
R> ind            <- which(y != 0)
R> workingMatrix  <- matrix(rep(y, length(y)), ncol = length(y))
\end{Sinput}
\end{Schunk}

The diagonal values of the matrix are reduced by one and only those columns of
the matrix with non-zero responses are kept.

\begin{Schunk}
\begin{Sinput}
R> diag(workingMatrix) <- diag(workingMatrix) - 1
R> workingMatrix       <- workingMatrix[, ind]
\end{Sinput}
\end{Schunk}

Now the \code{refit()} function can be applied to the columns of the matrix in 
order to obtain the estimates $\log \hat{\mu}_i(\bm{y}_{-i},y_i-1)$ in 
(\ref{eq:cAICpo}) from the reduced data.

\begin{Schunk}
\begin{Sinput}
R> workingEta  <- diag(apply(workingMatrix, 2, function(x) 
+   refit(p1, newresp = x)@resp$eta)[ind,])
\end{Sinput}
\end{Schunk}

The computation of the bias correction is then straightforward:

\begin{Schunk}
\begin{Sinput}
R> sum(y[ind] * (p1@resp$eta[ind] - workingEta))
\end{Sinput}
\end{Schunk}

\begin{Schunk}
\begin{Soutput}
[1] 205.5785
\end{Soutput}
\end{Schunk}

%
%
%
%
%
%

and corresponds to the bias correction obtained in Section~\ref{sec:cAIC_posBin}.

\subsection*{Bernoulli}

The computation of an estimator of the bias correction for Bernoulli distributed responses as in Equation~(\ref{eq:BerBC}) is 
similar to the implementation for Poisson distributed responses above. Therefore 
consider any Bernoulli model \code{b1} fitted by the \code{glmer} function in \pkg{lem4}. For the 
calculation of the bias correction for each observed response variable the 
model needs to be refitted with corresponding other value, i.e., $0$ for $1$ and vice versa. This is done best by use of the
\code{refit()} function from \pkg{lme4}.

\begin{Schunk}
\begin{Sinput}
R> muHat   	   <- b1@resp$mu
R> workingEta 	 <- numeric(length(muHat))
R> for(i in 1:length(muHat)){
+   workingData    <- b1$y
+   workingData[i] <- 1 - workingData[i]
+   workingModel   <- refit(b1, nresp = workingData)
+   workingEta[i]  <- log(workingModel@resp$mu[i] / 
+ 	(1 - workingModel@resp$mu[i])) - 
+ 	log(muHat[i] / (1 - muHat[i]))
+ }
\end{Sinput}
\end{Schunk}

The sign of the re-estimated logit (the natural parameter) in (\ref{eq:BerBC}) 
which is stored in the vector \code{workingEta} needs to be taken into 
account, i.e., $\hat{\eta}_i(1)$ is positive and $\hat{\eta}_i(0)$ negative. 
With a simple sign correction

\begin{Schunk}
\begin{Sinput}
R> signCor <- - 2 * b1@resp$y + 1
\end{Sinput}
\end{Schunk}

the following returns the bias correction:

\begin{Schunk}
\begin{Sinput}
R> sum(muHat * (1 - muHat) * signCor * workingEta)
\end{Sinput}
\end{Schunk}

It should be pointed out that for the conditional AIC it is essential to use the 
conditional log-likelihood with the appropriate bias correction. Notice that
the log-likelihood that by default is calculated by the S3-method
\code{logLik} for class \code{merMod} (the class of a mixed model fitted by 
a \code{lmer} call) is the marginal log-likelihood.

\bibliography{mybibliography}

\end{document}